\documentclass[12pt,preprint]{aastex}

\begin{document}

\title{Orbital Migration of Protoplanets in a Marginally Gravitationally Unstable Disk. II. Migration, Merging, and Ejection}

\author{Alan P. Boss}
\affil{Earth \& Planets Laboratory, Carnegie Institution
for Science, 5241 Broad Branch Road, NW, Washington, DC 20015-1305}
\authoremail{aboss@carnegiescience.edu}

\begin{abstract}

 Protoplanets formed in a marginally gravitationally unstable (MGU) disk by either core accretion or disk instability will
be subject to dynamical interactions with massive spiral arms, possibly resulting in inward or outward
orbital migration, mergers with each other, or even outright ejection from the protoplanetary system. The latter
process has been hypothesized as a possible formation scenario for the unexpectedly high frequency of
unbound gas giant exoplanets (free floating planets, FFP). Previous calculations with the EDTONS fixed grid
three dimensional (3D) hydrodynamics code found that protoplanets with masses from 0.01 $M_\oplus$ to 3 $M_{Jup}$
could undergo chaotic orbital evolutions in MGU disks for $\sim$ 1000 yrs without undergoing monotonic inward or
outward migration. Here the Enzo 2.5 adaptive mesh refinement (AMR) 3D hydrodynamics code
is used to follow the formation and orbital evolution of protoplanets in MGU disks for up to 2000 yrs. 
The Enzo results confirm the basic disk fragmentation results of the EDTONS code, as well as the absence of
monotonic inward or outward orbital migration. In addition, Enzo allows protoplanet mergers to occur, unlike 
EDTONS, resulting in a significant decrease in the number of protoplanets that survive for 1000 to 2000 yrs in the 
Enzo models. These models also imply that gas giants should be ejected frequently in MGU disks that fragment
into large numbers of protoplanets, supporting ejection as a possible source mechanism for the observed FFPs. 
 
\end{abstract}


\section{Introduction}

 Exoplanet demographics provide one of the ultimate arbiters of theories of exoplanet formation and evolution.
Nielsen et al. (2019) used the GPI exoplanet survey to search for planets with masses between
2 and 13 $M_{Jup}$ and semimajor axes between 3 and 100 au, finding that the peak occurrence distance
of giant planets was in the range of 1 to 10 au. Fulton et al. (2021) found the same peak occurrence 
distance of 1 to 10 au for the California Legacy Doppler velocities survey. Vigan et al. (2021) 
showed that the VLT SPHERES direct imaging survey of 150 stars detected 13 sub-stellar
companions with masses between 1 and 75 $M_{Jup}$ and semimajor axes between 5 and 300 au,
finding that both core accretion (CA; Mizuno 1980) and disk instability (DI; Boss 1997) 
appeared necessary to explain the detections for the FGK stars in their sample. 

 Gas giant planets with orbital distances as large as 980 au have been discovered and studied (Wu et al. 2022).
Forming giant exoplanets at such large distances by CA within the $\sim$ 1 Myr lifetimes of the  gaseous 
portion of protoplanetary disks is challenging (e.g., Chambers 2021), if not impossible.
DI has the advantage of forming dense, self-gravitating clumps in a few orbital periods, relaxing the disk 
lifetime constraint for forming wide-orbit gas giants in situ (e.g., Boss 2011).
Evidence for a gas giant protoplanet embedded in a spiral arm 93 au from AB Aurigae has been
interpreted as an example of gas giant planet formation by DI (Currie et al. 2022; cf. Cadman et al. 2021;
Zhou et al. 2022). DI has also been proposed as the source of the  $\sim 10 M_{Jup}$ exoplanet that 
orbits $\sim$ 560 au from the massive binary b Centauri (Janson et al. 2021).
Goda \& Matsuo (2019) examined the demographics of 485 planetary systems and concluded
that a hybrid theory of planet formation, involving both CA and DI, was needed to explain the
exoplanet detections.

 Miret-Roig et al. (2022) used a direct imaging survey coupled with Gaia and Hipparchos astrometry to
search for unbound gas giant exoplanets in the Upper Scorpius and Ophiuchus young stellar association.
Their survey yielded between 70 and 170 free floating planets (FFP), considerably more than might be
expected to form as  the tail end of the star formation process of molecular cloud core collapse and
fragmentation, and suggested that ejection from unstable planetary systems might make a major contribution
during the first 10 Myr. Gravitational microlensing has also found an abundance of likely FFPs, though 
these could also simply be bound planets with orbital distances greater than about 10 au 
(Mr\'oz et al. 2020; Ryu et al. 2021). Vorobyov (2016) performed numerical simulations that
supported the hypothesis that FFPs might be the result of planets ejected from massive MGU disks.

 While exoplanet demographics reveal orbital characteristics at the present epoch, unless exoplanets
do not undergo significant orbital evolution or migration following their formation, the present epoch 
orbital parameters are of limited usefulness in constraining their initial orbital distances. CA is the favored 
mechanism closer to the host star, as a result of shorter orbital periods, higher gas disk temperatures,
and higher surface densities of solids, to name a few factors, while DI may be more effective at larger 
distances in suitably massive and cool protoplanetary disks. For either CA or DI, a key question then 
becomes the extent to which protoplanets might migrate away from their birth orbits to their present 
epoch orbits.

  As noted by Boss (2013), CA and DI both require giant protoplanets to form in the presence of disk gas. 
Theoretical work on protoplanetary orbital migration (e.g., Kley \& Nelson 2012) usually focuses on 
protoplanets in disks where the disk mass is low enough that the disk self-gravity can be neglected, greatly
simplifying the analysis. Protoplanet evolution in MGU disk models has been calculated by Boss (2005), 
Baruteau et al. (2011), and Michael et al. (2011). These studies each considered quite different initial 
conditions and found a wide range of outcomes, ranging from large-scale inward orbital migration to 
relatively little orbital migration. Boss (2013) studied the evolution of protoplanets formed by either
CA and DI in MGU disks, noting that while a MGU disk is essential for formation by DI, even a giant planet
formed by CA in a quiescent, non-MGU disk can experience a later phase of MGU disk interactions during the 
periodic FU Orionis outbursts experienced by young solar-type protostars, which are thought to involve a phase of 
disk gravitational instability that dumps disk mass onto the protostar (e.g., Zhu et al. 2010; Kuffmeier et al. 2018). 
Dunhill (2018) similarly suggested that giant planets formed by CA might undergo orbital migration during 
FU Orionis outbursts. 

 The Boss (2005, 2013) models were performed using the EDTONS three dimensional radiative hydrodynamics code, 
with a spherical coordinate grid that was fixed at moderate spatial resolution throughout the MGU disk evolutions.
Virtual protoplanets were introduced at the beginning of each model to represent protoplanets as point
sources of gravity, able to interact gravitationally with the disk and with each other and to accrete mass from the disk by
Bondi-Hoyle accretion. Boss (2013) found that protoplanets with initial masses in the range from 
0.01 $M_\oplus$ to 3 $M_{Jup}$ and initial orbital distances of 6 to 12 au in a MGU disk 
around a solar-mass protostar underwent chaotic orbital evolutions for $\sim$ 1000 yr 
without undergoing the monotonic inward or outward migration that typically characterizes the Type I or 
Type II behavior of non-self-gravitating disk models (e.g., Kley \& Nelson 2012).

 The present models of protoplanet orbital evolution employ the Enzo 2.5 hydrodynamics code. 
Enzo is also a three dimensional (3D) code and uses Adaptive Mesh Refinement (AMR) in Cartesian coordinates
to ensure that sharp gradients in fluid quantities such as shock fronts can be handled accurately. 
Enzo is able to replace exceptionally dense disk clumps with sink particles representing newly formed (by DI)
protoplanets, which thereafter interact with each other and the disk while accreting disk gas, as do
the virtual protoplanets in the Boss (2013) models. We thus seek here to use a completely different 3D hydro code to
learn more about the possible outcomes for protoplanet orbital evolution in MGU disks, and to compare
the results with the latest advances in exoplanet demographics.
 
\section{Numerical Hydrodynamics Code}

 As noted by Boss \& Keiser (2013), the Enzo 2.5 AMR code performs hydrodynamics (HD) using any one of three 
different algorithms (Collins et al. 2010; Bryan et al. 2014): 
(1) the piecewise parabolic method (PPM) of Colella \& Woodward (1984), (2) the ZEUS method of Stone \& Norman 
(1992), or (3) a Runge–Kutta third-order-based MUSCL (“monotone upstream-centered schemes for conservation laws”) 
algorithm based on the Godunov (1959) shock-handling HD method. Enzo is designed for handling strong shock fronts 
by solving the Riemann problem (e.g., Godunov 1959) for discontinuous solutions of a fluid quantity that should be 
conserved. The PPM option was used in the current models as a result of the testing on mass and angular momentum 
conservation performed with Enzo 2.0 by Boss \& Keiser (2013), who found that PPM was better able to conserve mass and
angular momentum during the collapse of a rotating isothermal cloud core (Boss \& Bodenheimer 1979) than
either ZEUS or MUSCL. Enzo is designed for parallel processing on high performance clusters (HPC), but when run on
a single, dedicated 32-core node of the Carnegie memex HPC, a typical model still required 7 months of continuous
computation to evolve for $\sim 10^3$ yrs of model time.

  The Enzo 2.5 models were initialized on a 3D Cartesian grid with 32 top grid points in each direction. 
A maximum of 7 levels of refinement was used, with a factor of two refinement occurring for each level, so 
that the maximum possible effective grid resolution was $2^7$ = 128 times higher than the initial resolution 
of $32^3$, i.e., $4096^3$. The models with 7 levels needed an increase in the number of cell buffer zones
(NumberBufferZones) to 3 from the default value of 1, which was used for the lower levels of refinement,
in order to maintain reasonable time steps. Grid refinement was performed whenever necessary to ensure that 
the Jeans length constraint (e.g., Truelove et al. 1997; Boss et al. 2000) was satisfied by a factor of 4 for cells
with a density at least eight times the initial density. Periodic boundary conditions 
were applied on each face of the grid cubic box, with each side either 60 au or 120 au in length.
A  point source of external gravity was used to represent a 1 $M_\oplus$ protostar at the center of the grid.
The maximum number of Green’s functions used to calculate the gravitational potential was 10. The 
time step typically used was 0.15 of the limiting Courant time step. The results were analyzed with the 
yt astrophysical analysis and visualization toolkit (Turk et al. 2011).

 Following Boss \& Keiser (2014), we used the Enzo 2.2 sink particle coding described by Wang et al. (2010). Sink particles are 
created in grid cells that have already been refined to the maximum extent permitted by the specified number of levels 
of grid refinement, but where the gas density still exceeds that consistent with the Jeans length criterion for avoiding
spurious fragmentation (Truelove et al. 1997; Boss et al. 2000). As described by Boss \& Keiser (2014),
sink particles accrete gas from their host cells at the modified Bondi-Hoyle accretion rate proposed by Ruffert (1994). 
Two parameters control the conditions under which sink particles can be merged together: the merging mass (SinkMergeMass) 
and the merging distance (SinkMergeDistance). The former of these two parameters is used to divide 
the sink particles into either large or small particles. Particles with less mass than SinkMergeMass are first subjected 
to being combined with any large particles that are located within the SinkMergeDistance. Any surviving small particles 
after this first step are then merged with any other small particles within the SinkMergeDistance. The merging process 
is performed in such a way as to ensure conservation of mass and linear momentum. Boss \& Keiser (2014) found that
their results for collapse and fragmentation of magnetic molecular cloud cores were not particularly sensitive
to the choice of these two key parameters with regard to the tendency of the cores to undergo fragmentation into
multiple protostar systems. The current paper uses the Wang et al. (2010) sink particle coding with the 
SinkMergeMass set equal to 0.01 $M_{Jup}$ and the SinkMergeDistance set equal to 0.1 au, appropriate
values for studying gas giant protoplanets in a 120 au-size region. Sink creation was only allowed for cells with
densities exceeding the values listed in Table 1 (DensThresh in code units in the $sink\_maker.C$ subroutine).
These densities were chosen to be low enough that sinks do form in the models, as the point of the present
models was to study the orbital evolution of sink particles representing protoplanets in MGU disks rather than 
to study the precise physics of DI-induced fragmentation and clump formation in such disks 
(e.g., Boss 2021a,b). 

 The sink particles used in the Enzo models are similar to the virtual protoplanets (VPs) 
used in the EDTONS models: both are introduced in regions of density maxima and are intended to represent
gravitationally bound clumps of disk gas that will contract to form gaseous protoplanets, as they orbit in the
disk around the central protostar, interacting gravitationally with each other and the disk gas, even as they accrete 
more disk gas. There are several differences, however. Sink particles are created automatically by Enzo following
the criteria noted above, sink particles with close encounters can be merged together, and sink particles that encounter a 
grid boundary reappear on the opposite boundary as a result of the periodic boundary conditions. VPs, on the other 
hand, are inserted when a density maximum exceeds the Jeans length or Toomre length criteria 
(Nelson 2006; Boss 2021a,b) for the current grid spatial resolution. VPs may undergo close encounters with
each other but do not suffer mergers. VPs that strike either the inner or outer grid boundary are removed from
the calculation.

 While it would be desirable to compare flux-limited diffusion (FLD) approximation radiative 
hydrodynamic models from the EDTONS code with FLD radiative hydrodynamic models calculated by 
Enzo, the FLD routines available in Enzo are limited to non-local thermodynamic equilibrium (non-LTE), 
as Enzo was developed primarily for cosmological simulations, whereas EDTONS assumes LTE. As a 
result, we are limited to using a simpler approach to handling the disk thermodynamics with the Enzo code.
Boss (1998) showed that disk fragmentation could occur for strongly gravitationally unstable
disks with either locally isothermal or locally adiabatic thermodynamics, using disk gas adiabatic 
exponents ranging from $\gamma = 1$ (purely isothermal) to $\gamma = 7/5$, which is appropriate 
for molecular hydrogen. Given that disks are subject to compressional heating,  $\gamma = 1$ is
not strictly correct, and given that disks that are optically thick at their midplanes can cool from their 
surfaces,  $\gamma = 7/5$ is not strictly correct either. The physically correct behavior presumably
lies somewhere in the middle of these two extremes. 

 Radiative cooling in optically thin regions was employed in the Enzo models, with a critical density 
for cooling of  $10^{-13}$ g cm$^{-3}$; regions with densities above this critical value had the cooling 
rate decreased proportionally. This critical density was chosen because that is the disk midplane
density where the dust grain opacity produces optical depths of order unity (e.g., Boss 1986).
The cooling rates were modified from the default values in $cool\_rates.in$ to rates consistent
with molecular line cooling in optically thin regions (Boss et al. 2010; Neufeld \& Kaufman 1993).
Because Enzo PPM hydrodynamics involves a Riemann solver that cannot be purely isothermal,
i.e., $\gamma$ cannot equal unity, the adiabatic index for the disk gas was taken to be 
$\gamma = 1.001$, appropriate for a nearly isothermal, but still adiabatic equation of state for 
an ideal gas. Test runs were computed for 100 yrs of evolution with both $\gamma = 7/5$
and $\gamma = 5/3$, but in both cases Enzo produced midplane disk temperatures that were 
over $10^4$ K, whereas the initial disk had a maximum midplane temperature of 1500 K. 
The test runs with $\gamma = 1.001$ produced the expected maximum temperatures of
$\sim 1500$ K, and hence $\gamma = 1.001$ was adopted for the models presented here.
The resulting temperature distributions were also affected by 
the assumption of radiative cooling; spiral features in the midplane temperature distribution 
accompanied spiral features in the midplane density distribution, as is to be expected.
Finally, the mean molecular weight of the gas was effectively taken to be $\mu = 2.4$, 
appropriate for a solar composition mixture of molecular hydrogen and helium.

\section{Initial Conditions}

 Table 1 lists the models with variations in the number of levels of grid refinement, the outer disk 
and envelope temperatures, initial minimum value of the Toomre (1964) $Q$ parameter,
disk radius, calculational grid box size, and critical density for sink
particle creation. A 60 au box size was used for the 20 au and 30 au radius disks, while a 120 au box size 
was used for 60 au radius disks, in order to give the disks sufficient room to evolve and expand
by the outward transport of angular momentum through gravitational interactions with the
spiral arms and clumps. In the number of levels column, 34 means the model was initially
run with 3 levels and then a fourth level of refinement was added.

 The initial disks are based on the model HR disk from Boss (2001), with an outer disk temperature of 
40 K and and disk envelope temperature of 50 K, which has been used as a standard initial
model for many of the author's disk instability models (e.g., Boss 2021a,b).
Model HR has an initial minimum Toomre $Q \approx 1.3$, implying marginal stability to the
growth of rings and spiral arms. The model HR initial disk has a mass of 0.091
$M_\odot$ within an inner radius of 4 au and an outer radius of 20 au and 
orbits a 1 $M_\odot$ central protostar. The Enzo models have have masses of 
0.102 $M_\odot$ for 20 au outer radius disks, slightly higher than in model HR because 
the Enzo models extend inward to 1 au, 0.142 $M_\odot$ for 30 au 
outer radius disks, and 0.210 $M_\odot$ for 60 au outer radius disks.
The same disk density power-law-like Keplerian structure as in Boss (2001) is used for all
of the models, with the structure being terminated at 20 au, 30 au, or 60 au. 
Figures 1 and 2 show cross sections of the initial disk density distribution for the 20 au disks,
both parallel and perpendicular (i.e., disk midplane) to the disk rotation axis.

\section{Results}

 Figure 3 shows the intermediate results for two of the four models that have
the identical initial disk configuration (20 au radius) as the Boss (2001) model HR,
depicted at the same time (190 yrs of evolution) as the same initial disk
model (fldA) in Boss (2021b, cf. Figure 2a). Figure 3 shows that both of these
models (3-1K-20 and 6-1K-20) rapidly evolved into a configuration of multiple 
spiral arms interspersed with dense clumps, as expected for a marginally 
gravitationally unstable disk. Also as expected, the degree of fragmentation
and clump formation increases as the numerical grid resolution increases from
3 to 6 levels. When sink particles are allowed to form, the number of sink particles
similarly increases as the resolution is improved. While the background disk
looks quite similar for model 3-1K-20 with or without sink particles (Figure 3a,c),
there is a clear difference in the case of model 6-1K-20 (Figure 3b,d), where
the background disk has become perturbed into a prolate configuration due 
to the formation of a massive ($\sim 20 M_{Jup}$) secondary companion 
(at one o'clock), with its own circumplanetary disk and tertiary companion, 
whose combined tidal forces have evidently distorted the disk's overall appearance.
Model fldA of Boss (2021b) had fragmented into a five clumps and three virtual
protoplanets (i.e., sink particles) by 189 yrs, for a total of eight, considerably
more than formed in the present model 3-1K-20, but not as many as in
model 6-1K-20, suggesting that even with the quadrupled spatial resolution
of the Boss (2021b) EDTONS models, the adaptive mesh refinement feature of
Enzo results in significantly improved numerical spatial resolution of the 
disk instability and fragmentation. Confirmation of the formation of long-lived
fragments in the model HR disk (Boss 2001, 2021b) with the completely different 
hydrodynamical method used here provides strong support for the viability
of the disk instability mechanism for the formation of gas giant protoplanets
and higher mass companions.

 Figure 4 displays the results after 2000 yrs for the Enzo models in Figure 3. The 
EDTONS model fldA in Boss (2021b) was stopped after only 189 yrs of evolution, 
but even still required over 4 years of computation on a single core of a node on 
memex, a Carnegie Institution computer cluster. EDTONS is based on code initially
written in the late 1970s and is not parallelized. Enzo, in contrast, is a modern
code designed to run on parallel processing systems like memex, and as a
result the Enzo models can be computed much farther in time. Even still,
model 3-1K-20 required one week to run for 2000 yrs of model time on a dedicated 
single memex node with 28 cores, while model 6-1K-20 required 
one year to run 2000 yrs on a dedicated 28-core node. Three dimensional
hydrodynamics at high spatial resolution is computationally expensive,
even when a parallelized code is employed.

 Figure 4 shows that the evolution of these two models diverged considerably
following the early fragmentation phase depicted in Figure 3. The two sink
particles evident in Figure 4a have masses of $\sim 2 M_{Jup}$ and 
$\sim 0.6 M_{Jup}$, with a total gas disk mass of $\sim 99 M_{Jup}$,
while the 13-odd sink particles in Figure 4b have masses
ranging from $\sim 0.2 M_{Jup}$ to $\sim 23 M_{Jup}$, for a total sink particle
mass of $\sim 96 M_{Jup}$, leaving a disk gas mass of only $\sim 5 M_{Jup}$.
Clearly, the final disk mass in model 3-1K-20 far outweighs the mass of
the sink particles, and as a result the particles are unable to open gaps
in the disk, though the disk has expanded outward to a radius of about 30 au
as a result of the transport of disk mass and angular momentum outward,
caused by the strong spiral arms evident in Figure 3a,c. The fact that sink
particle formation has been so efficient in model 6-1K-20, with the total particle
mass some 20 times larger than the disk mass, means that the particles
rule the evolution and are able to clear out a distinct inner gap, centered
on about 5 au (Figure 4b). In model 6-1K-20, the sink particles gained the
bulk of the disk's mass and angular momentum, so that the disk is not
able to expand beyond its initial radius of 20 au. Three sink particles were
accelerated to speeds high enough at their orbital location to be ejected
altogether from the system, but because of the periodic boundary conditions
imposed on the calculations by the Enzo self-gravity solver, these ejectable
particles were returned to the system and underwent further interactions with the
sink particles and disk gas.

 Table 2 gives the maximum number of sink particles formed for all the 
models, as well as the number surviving at the end of the run. 
Table 2 shows that the maximum number of sinks formed decreases as
the initial disk gas temperature is increased, as this results in an increase
in the Toomre minimum $Q$ value (Table 1), i.e., in greater stability to the 
growth of rings and spiral arms, and hence to fragmentation and sink
particle formation. By the time that the disk temperature is increased to 160 K,
disk fragmentation is completely stifled in the Enzo models, consistent
with the flux-limited diffusion approximation models of Boss (2021b), 
where fragmentation ceased for a minimum Toomre $Q$ greater than 2.2.

 Models 6-2K-30 and 7-2K-30 did not form sink particles, unlike the 
otherwise identical, but lower resolutions models 3-2K-30, 4-2K-30, and 5-2K-30,
because this sequence used a fixed critical density for sink particle formation
of $10^{-9}$ g cm$^{-3}$. That choice meant that the dense clumps formed
in the two higher resolution models could always be resolved with more
grid levels and finer spatial resolution, thereby preventing the clumps from
exceeding the critical density required for sink particle formation, at least
during the limited amount of model time that the 6- and 7-level models
were able to be evolved (326 and 340 yrs, respectively). Small time steps
prevented these two models for being evolved farther in time. Figure 5 shows
these two models at their final times, showing that the spiral arms and
nascent clumps become more distinct as the number of grid levels is
increased, as expected when approaching the continuum limit of infinite
spatial resolution.

 Table 2 also lists the number of sink particles mergers,  where 
$N_{merged-sinks} = N_{max-sinks} - N_{final-sinks}$, and the number of
times that a sink particle would have been ejected if periodic boundary 
conditions were not required. Table 2 shows that mergers of sink
particles are quite common in all of the models that formed sink particles,
and evidently are responsible for much of the gain in mass of the particles, 
along with the ongoing accretion of disk gas, given that the number of
mergers is usually comparable to, or far greater than, the final number of
sink particles. The value $N_{escaped-sinks}$ can be quite large 
due to the sink particles' inability to escape the system; often the same
particle bounces in and out in orbital radius and achieves escape velocity
multiple times. Achieving the escape velocity usually occurs for particle
orbital distances of 30 au to 40 au, but can also occur from 10 au to 20 au in
the more unstable disks (e.g., 5-1K-20, 6-1K-20). The large numbers of
escape episodes in the latter two models are clearly solely a result
of the periodic boundary conditions, but they do indicate that 
ejected protoplanets are to be expected as a natural outcome of
a phase of gas disk gravitational instability. Table 2 suggests that
such a phase of protoplanetary disk evolution should result in the
ejection of several gas giant protoplanets.

 Figure 6 presents all of the sink particle masses and distances from the
central star at the final times for the models. These distances correspond 
to observed separations in the absence of any other knowledge of the orbital 
parameters, i.e., the semimajor axis and eccentricity, The final masses
range from $\sim 0.1 M_{Jup}$ to $\sim 100 M_{Jup}$, i.e., sub-Jupiters
to brown dwarfs and late M dwarf stars. Separations range from inside
1 au to over 30 au. Ejected particles would be at much larger distances,
were ejection permitted. 

 Figure 6 shows that the black dots, representing
the 20 au radius disks, tend to have higher masses ($> 10 M_{Jup}$)
inside 10 au than the blue dots, representing the 60 au radius disks,
which tend to have lower masses ($< 1 M_{Jup}$) inside 10 au.
This outcome is the result of the 20 au radius disks all starting
their evolutions from considerably more gravitationally unstable
initial states, i.e., Toomre $Q_{minimum} = 1.3$ than the 60 au radius
disks, with initial Toomre $Q_{minimum} = 1.9$ or 2.2. The 20 au radius
models thus generally form more massive sink particles, as would be
expected.

 Figure 7 shows the sink particle masses as a function of the orbital 
semimajor axis at the final times for the models, while Figure 8
depicts these properties for the known exoplanets on the same
scales. Figure 3b shows that fragmenting dense clumps appear between
about 5 au and 20 au, which is the same distance range as most of the
sink particles in Figure 7; only a few have migrated inside 1 au,
and only a few orbit beyond about 20 au.
Clearly the present models produce a goodly number of
cool gas giants and brown dwarfs, but do not lend support for the
formation and inward migration of the numerous hot and warm exoplanets
evident in Figure 8: little evidence for monotonic inward orbital
migration is seen. This result is consistent with the EDTONS models
of Boss (2013).

 Finally, Figure 9 shows the sink particle masses as a function of the orbital 
eccentricity at the final times for the models, while Figure 10
depicts these properties for the known exoplanets on the same
scales. The present models show that the processes studied here
of fragmentation, mergers, chaotic orbits, and ejections result in
the observed wide range of eccentricities, though not the 
presumably tidally damped, near-zero eccentricities of the hot Jupiters.

\section{Discussion}

 Drass et al. (2016)  showed that the initial mass function in the Orion nebula cloud has two peaks,
one at 0.25 $M_\odot$ and another at 0.025 $M_\odot$, and suggested that the latter peak was
composed of brown dwarfs and isolated planetary-mass objects that had been ejected from circumstellar
disks or multiple star systems. The large number of attempted ejections in the Enzo models
that are listed in Table 2 fully support this hypothesis.

 Feng et al. (2022) combined high-precision Doppler velocity data with Gaia and Hipparcos astrometry 
to constrain the masses and orbital parameters of 167 sub-stellar companions to nearby stars. Their
Figure 3 shows that these 167 companions fully populate a parameter space ranging from semimajor
axes of $\sim$ 2 au to $\sim$ 40 au, with masses from $ \sim 4 M_{Jup}$ to  $ \sim 100 M_{Jup}$, 
much like the upper right quadrant of Figure 7. Their Figure 3 also shows orbital eccentricities varying
from 0 to 0.75, again in basic agreement with the range evident in the present models in Figure 9.
These Enzo models suggest a unified formation mechanism of the 167 sub-stellar companions
studied by Feng et al. (2022): fragmentation of MGU disks.

 Galvagni et al. (2012) used a smoothed particle hydrodynamics (SPH) code to study clumps
formed at $\sim$ 100 au in a MGU disk, finding that the clumps could contract and heat up
enough to begin molecular hydrogen dissociation, resulting in a dynamical collapse phase
that can ensure their survival to tidal forces. Their results showed that this collapse phase
could occur within $\sim 10^3$ yrs, shorter than the evolution times of the models considered 
here (Table 2), justifying the replacement of dense clumps with Enzo sink particles or  
EDTONS virtual protoplanets (e.g., Boss 2005, 2013).

 Lichtenberg \& Schleicher (2015) used Enzo to study fragments formed by the disk instability process in 
isothermal disks, but did not employ sink particles or radiative transfer effects, finding that the
clumps formed were all lost by inward migration combined with the tidal force of the protostar.
Stamatellos  (2015) used a SPH code to study disks with radii of 100 au and high Toomre $Q$
values. Planets inserted at 50 au either migrated inward or outward over $2 \times 10^4$ yrs, depending
on whether they were allowed to gain mass or not, respectively. 

 Hall et al. (2017) used an SPH code to study the identification and interactions of disk fragments
composed of clumps of SPH particles that formed from the fragmentation of a $0.25 M_\odot$ disk of 
radius 100 au around a $1 M_\odot$ protostar. Their models showed that fragment-fragment interactions
early in the evolutions led to scattering of fragments to larger semi-major axes, as large as 250 au, and
to increased eccentricities, as high as 0.7. While the periodic boundary conditions used in the present
models preclude an assessment of the final semi-major axes after close encounters, the fact that
the sink particle velocities were often sufficiently high to predict ejection from the system is consistent
with the Hall et al. (2017) results showing efficient scattering outward (cf., Table 2). The eccentricity
pumping found by Hall et al. (2017) is similarly consistent with that found in the present models (cf. Figure 9).

 Hall et al. (2017) also studied tidal downsizing and disruption of fragments that ventured too 
close to the tidal forces of the central protostar, finding that more clumps were destroyed by tidal
disruption than by disappearing in a merger event. Tidal downsizing was proposed by Nayakshin (2010,
2017) as a means for forming inner rocky worlds from gas giants formed in a disk instability,
following the formation of rocky inner cores by the sedimentation of dust grains and pebbles
to the center of the giant gaseous protoplanet (Boss 1997). Tidal downsizing remains as a creative
means to form inner rocky worlds as a result of a gravitationally unstable gas disk. The present
sink particle models, as well as the virtual protoplanet models  of the EDTONS code, do not allow
tidal downsizing to occur, though implicitly the loss of virtual protoplanets that hit the inner
disk boundary in EDTONS code calculations could be considered the equivalent of the loss of
gas giant protoplanets by tidal disruption. Modeling the interior structure and thermal evolution
of slowly contracting gas giant protoplanets is a future challenge for these types of models, and 
tidal disruption could result in the loss of sink particles that pass close to the central protostar, 
though it can be seen in Figure 6 that few sink particles passed inside 1 au.

 Fletcher et al. (2019) performed a code comparison study of the orbital migration of 
protoplanets inserted at 120 au in disks of 300 au radius, finding that protoplanets of 2 $M_{Jup}$  
migrated inward to $\sim$ 40 au to $\sim$ 60 au within $\sim 10^4$ yr. These code comparisons differ considerably from
the present models, as only single protoplanets were injected, the disks used a $\gamma = 7/5$ adiabatic
index, and the disks were gravitationally stable everywhere, with Toomre $Q \ge 2$. As a result, the
evolutions did not undergo the chaotic evolutions of the present models, where the MGU disk produces
strong spiral arms that interact with the numerous protoplanets that formed near the outset.

 Finally, Rowther \& Meru (2020) used a SPH code to study
planet survival in self-gravitating disks. They found that a fixed-mass planet with a range of
masses would migrate inward in the cool outer regions of their disks, but that this 
migration was halted once the planet reached the warm inner disk. In their models,
a single planet at a time is embedded in a disk with a mild spiral arm structure. Compared to
the multiple clumps, sink particles, and strong spiral arms that form and interact in the present 
models (e.g., Figure 3), it is clear that the Rowther \& Meru (2020) planets do not undergo
the chaotic orbital motions experienced by the Enzo models here (or the EDTONS models
of Boss 2013), which prevent monotonic orbital migration.

\section{Conclusions}

 The use of a completely different three dimensional hydrodynamical code (Enzo 2.5), with a completely
different method for handling nascent protoplanets (sink particles), has produced results in good 
agreement with those obtained by the EDTONS code and the virtual protoplanet method (Boss 2005).
Both codes agree that with high spatial resolution, the standard model HR disk (Boss 2001)
fragments rapidly into multiple dense clumps and strong spiral arms. Both codes agree that
when these clumps are replaced with particles that can accrete mass from the disk, the
particles grow in mass and can orbit chaotically for 1000 yrs to 2000 yrs without suffering
monotonic inward or outward orbital migration. In addition, the Enzo models show that the
protoplanets have a high probability of close encounters with each other, leading either to
mergers, or to being ejected from the protoplanetary system. Comparisons with the observational
data on exoplanet demographics and FFPs suggest that gas disk gravitational instabilities
have an important role to play in explaining the formation of sub-stellar companions with a
wide range of masses and orbital distances.

\acknowledgments

 I thank Sean Raymond for discussions about FFPs and Floyd Fayton for his invaluable assistance with 
the memex cluster. I also thank the reviewer for providing several suggestions for improving
the manuscript. The computations were performed on the Carnegie Institution memex computer
cluster (hpc.carnegiescience.edu) with the support of the Carnegie Scientific
Computing Committee. The computations were performed using the Enzo code
originally developed by the Laboratory for Computational Astrophysics at the University 
of California San Diego and now available at https://enzo-project.org/.

\clearpage
\begin{deluxetable}{lccccccc}
\tablecaption{Initial conditions for the models with varied maximum
number of refinement levels, initial outer disk and envelope temperatures (K), 
initial minimum Toomre $Q$, outer disk radii (au), box size (au), and 
critical density needed for sink particle formation (cgs). \label{tbl-1}}
\tablewidth{0pt}
\tablehead{\colhead{Model} 
& \colhead{$N_{levels}$}
& \colhead{$T_{disk}$} 
& \colhead{$T_{envelope}$} 
& \colhead{$Q_{minimum}$} 
& \colhead{$R_{disk}$}
& \colhead{$S_{box}$}
& \colhead{$\rho_{crit-sinks}$}}
\startdata

3-1K-20  & 3       & 40  &  40 &  1.3 &  20  & 60 & $10^{-8}$ \\

4-1K-20  & 4       & 40  &  40 &  1.3 & 20  & 60 & $10^{-8}$ \\

5-1K-20  & 5       & 40  &  40 &  1.3 & 20  & 60 & $10^{-7}$ \\

6-1K-20  &  6      & 40  &  40 &  1.3 & 20  & 60 & $10^{-7}$ \\

3-2K-30 &  3 &  80 &   80 & 1.9 & 30  & 60 & $10^{-9}$  \\

4-2K-30 & 4 &   80 &   80 & 1.9 & 30  & 60 & $10^{-9}$  \\

5-2K-30 & 5  &  80 &   80 & 1.9 & 30  & 60 & $10^{-9}$  \\

6-2K-30 & 6  &  80 &   80 & 1.9 & 30  & 60 & $10^{-9}$  \\

7-2K-30 & 7  &  80 &   80 & 1.9 & 30  & 60 & $10^{-9}$  \\

3-2K-60-11   & 3 & 80 &  120 & 1.9 & 60 & 120 & $10^{-11}$  \\

34-2K-60-11 & 3-4 & 80 &  120 & 1.9 & 60  & 120 & $10^{-11}$  \\

4-2K-60-10   & 4  & 80 &   120 & 1.9 &  60 & 120 &   $10^{-10}$ \\

4-2K-60-11   & 4  & 80 & 120 & 1.9 & 60  & 120   & $10^{-11}$    \\

3-3K-60-10    & 3  &  120 & 120 & 2.2 & 60  & 120 & $10^{-10}$   \\

3-3K-60-11   & 3  & 120 & 120 & 2.2 & 60   & 120  & $10^{-11}$   \\

34-3K-60-10  & 3-4 & 120 &  120  & 2.2 & 60  & 120 &  $10^{-10}$  \\

34-3K-60-11 & 3-4 & 120 &  120 & 2.2 & 60   & 120  & $10^{-11}$   \\

4-3K-60-10   & 4  & 120 &   120 & 2.2 &  60  & 120 &  $10^{-10}$   \\

4-3K-60-11   & 4 & 120 &  120 & 2.2 &  60    & 120 & $10^{-11}$  \\

3-4K-60-10    & 3  & 160 &   120 & 2.5 & 60 & 120 &  $10^{-10}$  \\

3-4K-60-11   & 3 & 160  &  120 & 2.5 & 60  & 120   & $10^{-11}$  \\

\enddata
\end{deluxetable}

\clearpage
\begin{deluxetable}{lccccc}
\tablecaption{Results for the models, showing the 
maximum number of sinks formed, final number of sinks, number of times a
sink reached escape velocity, number of sinks lost to mergers, and final time (yrs). \label{tbl-1}}
\tablewidth{0pt}
\tablehead{\colhead{Model} 
& \colhead{$N_{max-sinks}$}
& \colhead{$N_{final-sinks}$}
& \colhead{$N_{escaped-sinks}$}
& \colhead{$N_{merged-sinks}$}
& \colhead{$t_{final}$}}
\startdata

3-1K-20   & 4  & 2 & 0 & 2 & 2000 \\

4-1K-20   & 11 & 3 & 0 & 8 & 2000 \\

5-1K-20   & 23 & 11 & 130 & 12 & 2000 \\

6-1K-20   & 30 & 18 &  540 & 12 & 2000 \\

3-2K-30  & 13  & 2 &  6 & 11 & 2000 \\

4-2K-30  & 15  & 3 &  50 & 12 & 2000 \\

5-2K-30  & 19  & 5 & 110 & 14 & 2000 \\

6-2K-30  & 0  & 0 &  0& 0 & 326 \\

7-2K-30  & 0  & 0 &  0& 0 & 340 \\

3-2K-60-11 &  25  & 4  &  0 & 21 &1000  \\

34-2K-60-11 & 12  & 2 &  15 & 10 & 1000  \\

4-2K-60-10   &  0  & 0 &  0& 0 &130 \\

4-2K-60-11   & 0  &  0 &  0& 0 & 234  \\

3-3K-60-10   & 12 & 2 &  0 & 10& 1000   \\

3-3K-60-11    &11 & 2 &  0 &  9 & 1000  \\

34-3K-60-10 &    10 &  3 & 0 & 7 & 890   \\

34-3K-60-11  & 10 & 5 &  0 & 5 & 1000  \\

4-3K-60-10    &   0  &  0  &  0& 0& 250   \\

4-3K-60-11    &  0  & 0 &  0& 0 & 268  \\

3-4K-60-10    &  0  & 0 &  0& 0& 86   \\

3-4K-60-11   &   0  & 0 &  0& 0& 142  \\

\enddata
\end{deluxetable}

\begin{figure}
\vspace{-1.0in}
\includegraphics[scale=.60,angle=-90]{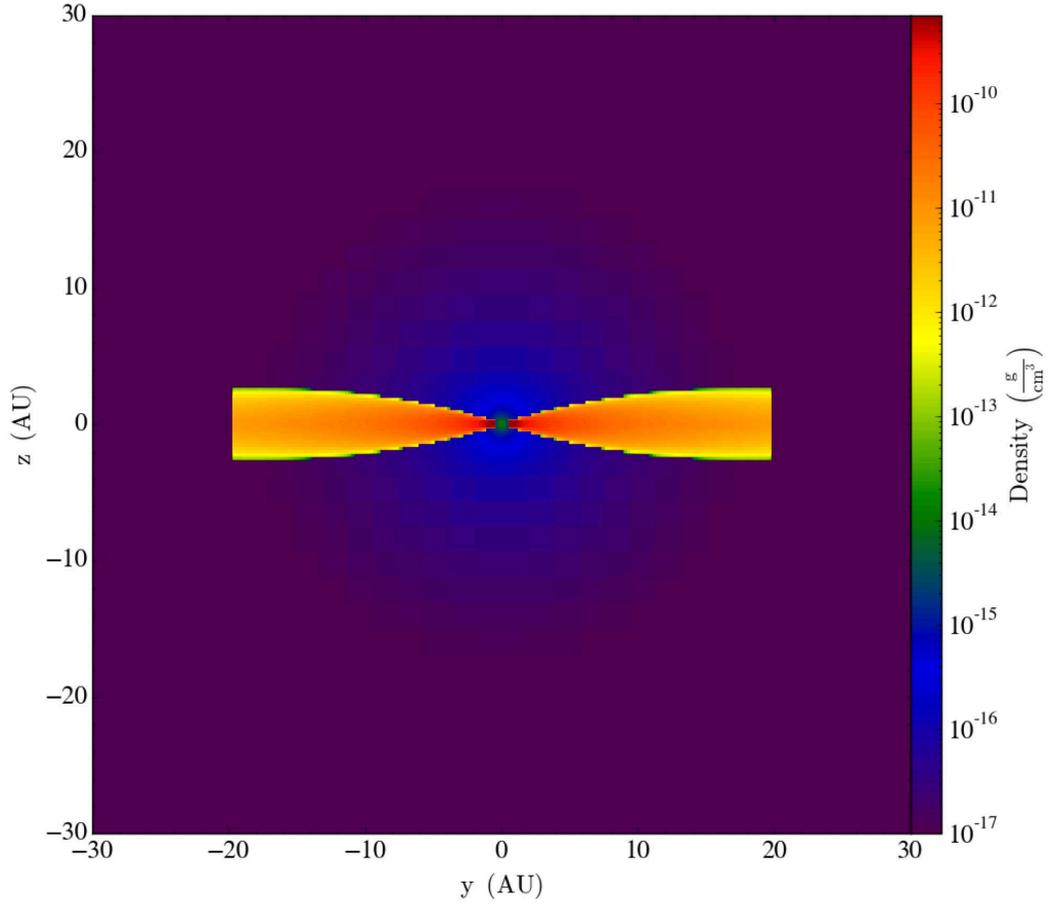}
\vspace{0.5in}
\caption{Initial log density cross-section in a vertical section ($x$ = 0) showing the entire computational 
grid with a maximum of three levels of refinement for the 20 au outer disk radius models. }
\end{figure}
\clearpage

\begin{figure}
\vspace{-1.0in}
\includegraphics[scale=.60,angle=-90]{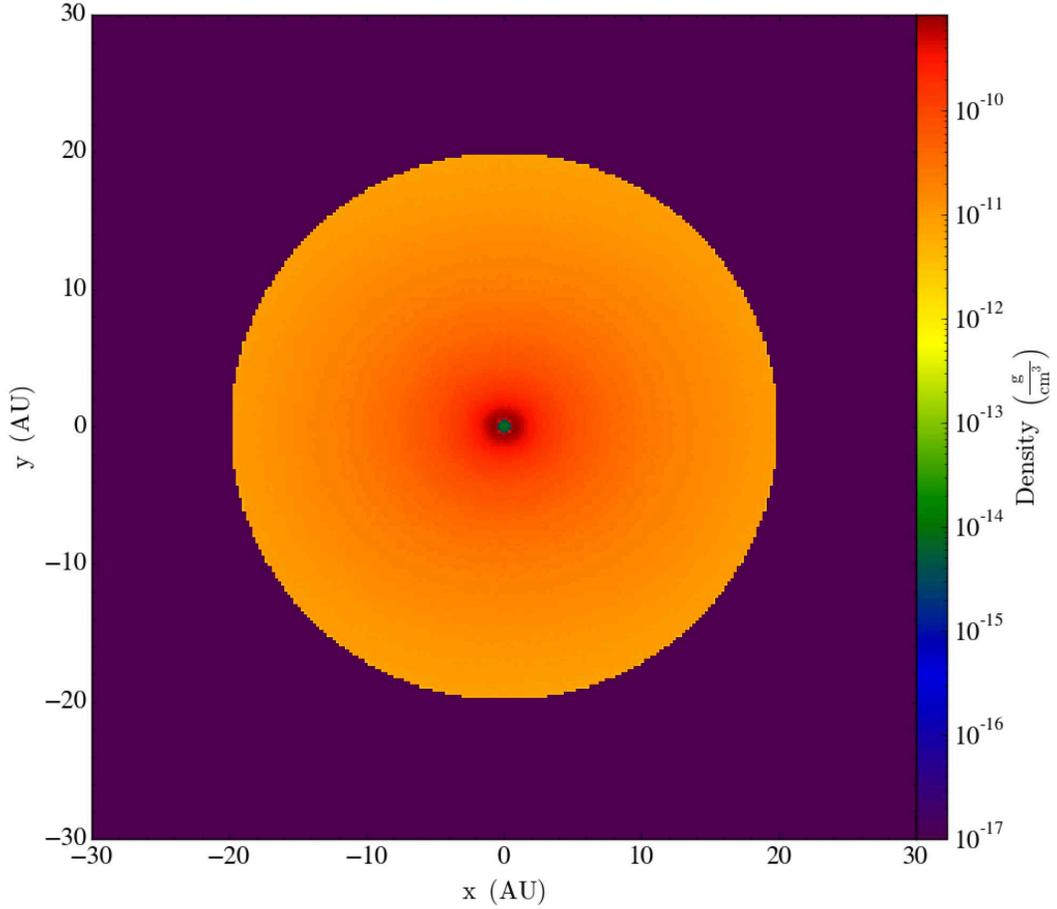}
\vspace{0.5in}
\caption{Initial log density cross-section in the disk midplane ($z$ = 0) showing the entire computational 
grid with a maximum of three levels of refinement for the 20 au outer disk radius models. With six
levels of refinement, the inner 1 au is better resolved, but otherwise the initial disk is identical. }
\end{figure}
\clearpage

\begin{figure}
\vspace{-1.0in}
\includegraphics[scale=0.8,angle=+90]{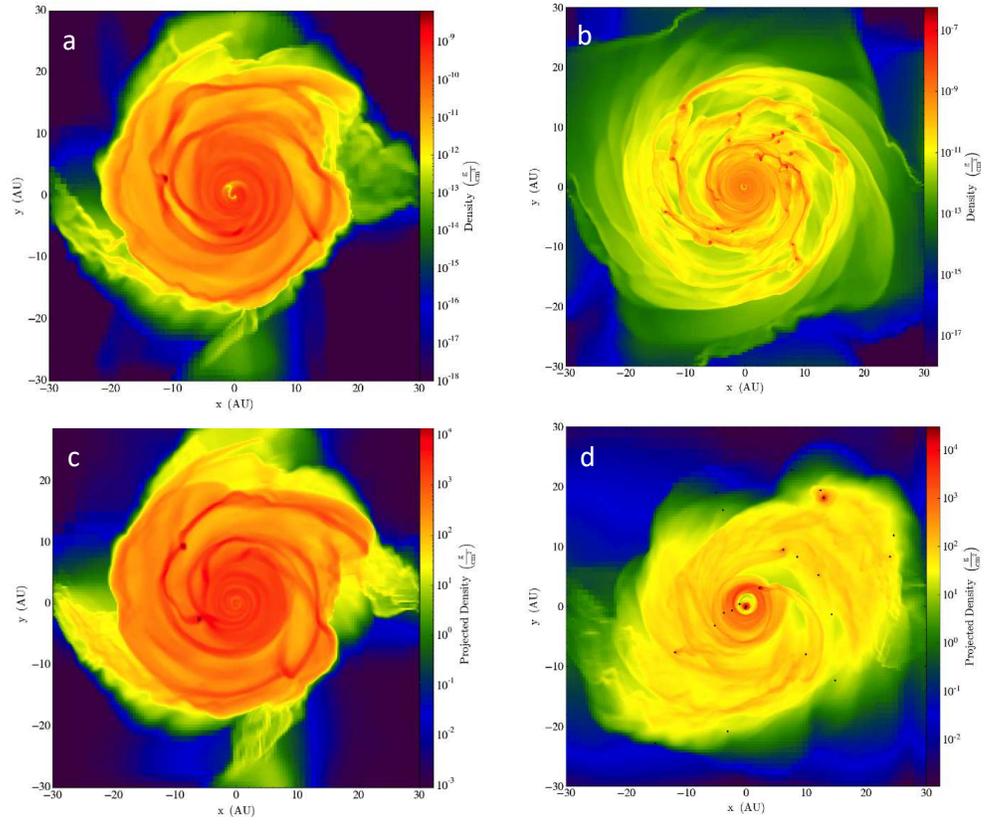}
\vspace{0.5in}
\caption{Log density cross-section in the disk midplane ($z$ = 0) after 190 yr of evolution for
model 3-1K-20 without (a) and with (c) sink particles, and 
model 6-1K-20 without (b) and with (d) sink particles. }
\end{figure}
\clearpage

\begin{figure}
\vspace{-1.0in}
\includegraphics[scale=0.8,angle=+90]{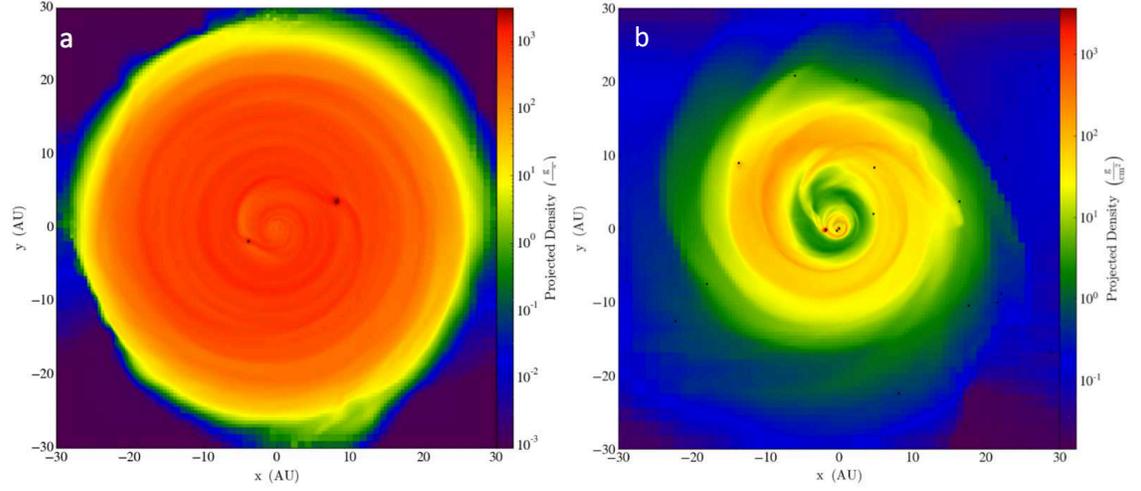}
\vspace{0.5in}
\caption{Log density cross-section in the disk midplane ($z$ = 0) after 2000 yr of evolution for
(a) model 3-1K-20 and (b) model 6-1K-20, both with sink particles. }
\end{figure}
\clearpage

\begin{figure}
\vspace{-1.0in}
\includegraphics[scale=0.8,angle=+90]{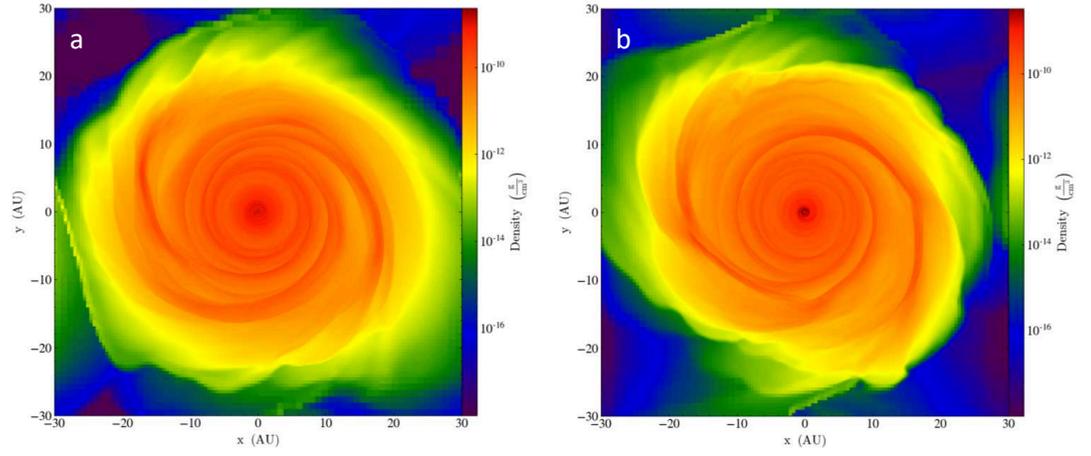}
\vspace{0.5in}
\caption{Log density cross-section in the disk midplane ($z$ = 0)  for
(a) model 6-2K-30 and (b) model 7-2K-30 after 326 yr and 340 yr of evolution, respectively. }
\end{figure}
\clearpage

\begin{figure}
\vspace{-1.0in}
\includegraphics[scale=.80,angle=-90]{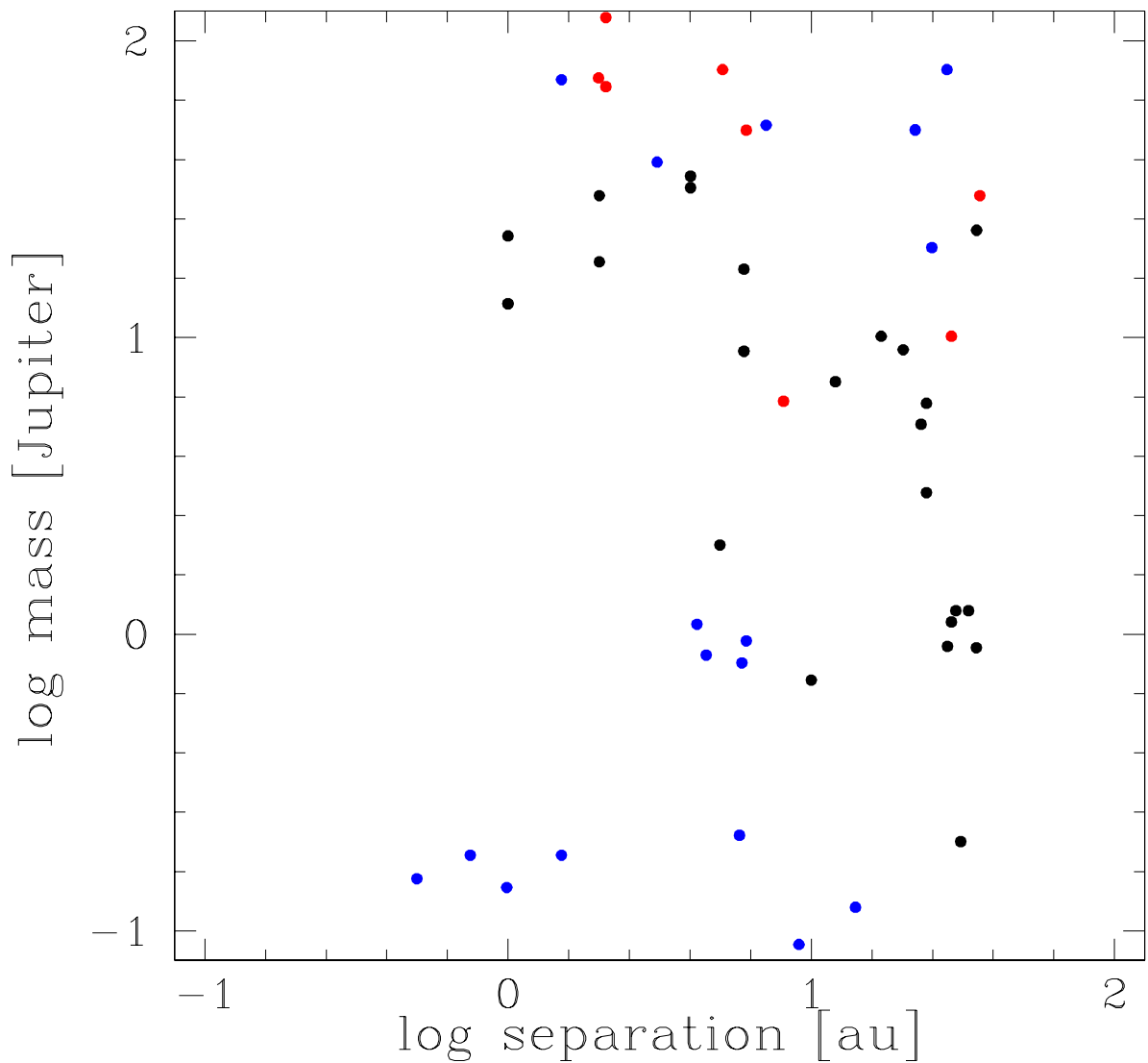}
\vspace{0.5in}
\caption{Sink particle masses and distances from central star at the final times for the 
models. These distances correspond to observed separations in the absence of knowledge
of the orbital parameters, i.e., the semi-major axis and eccentricity. Black dots are for
models that started with 20 au radius disks, red dots are for 30 au disks, and blue
dots are for 60 au radius disks (see Table 1).}
\end{figure}
\clearpage

\begin{figure}
\vspace{-1.0in}
\includegraphics[scale=.80,angle=-90]{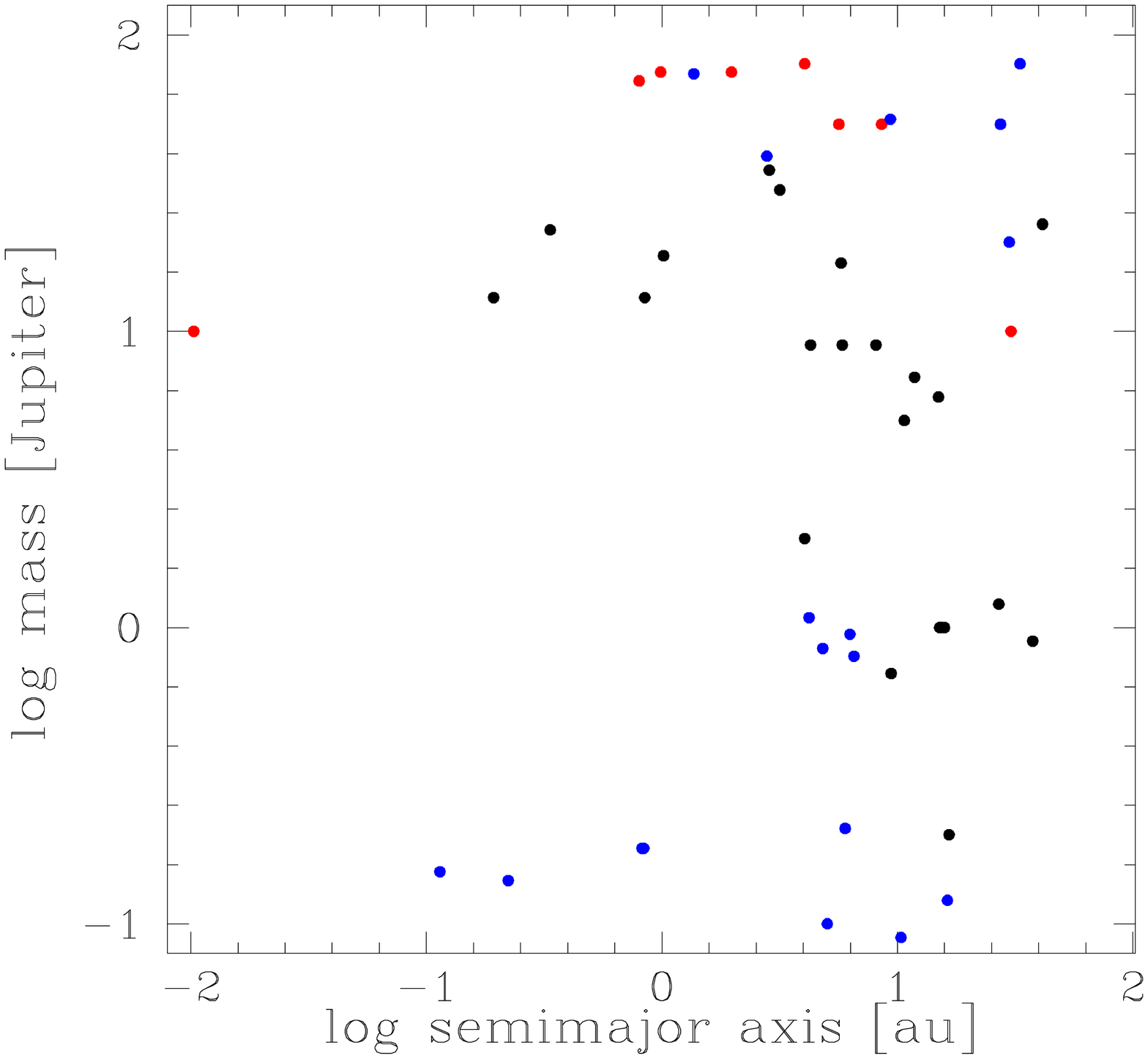}
\vspace{0.5in}
\caption{Sink particle masses as a function of the orbital semimajor axis at the final times 
for the models. Black dots are for
models that started with 20 au radius disks, red dots are for 30 au disks, and blue
dots are for 60 au radius disks (see Table 1).}
\end{figure}
\clearpage

\begin{figure}
\vspace{-1.0in}
\includegraphics[scale=.60,angle=-90]{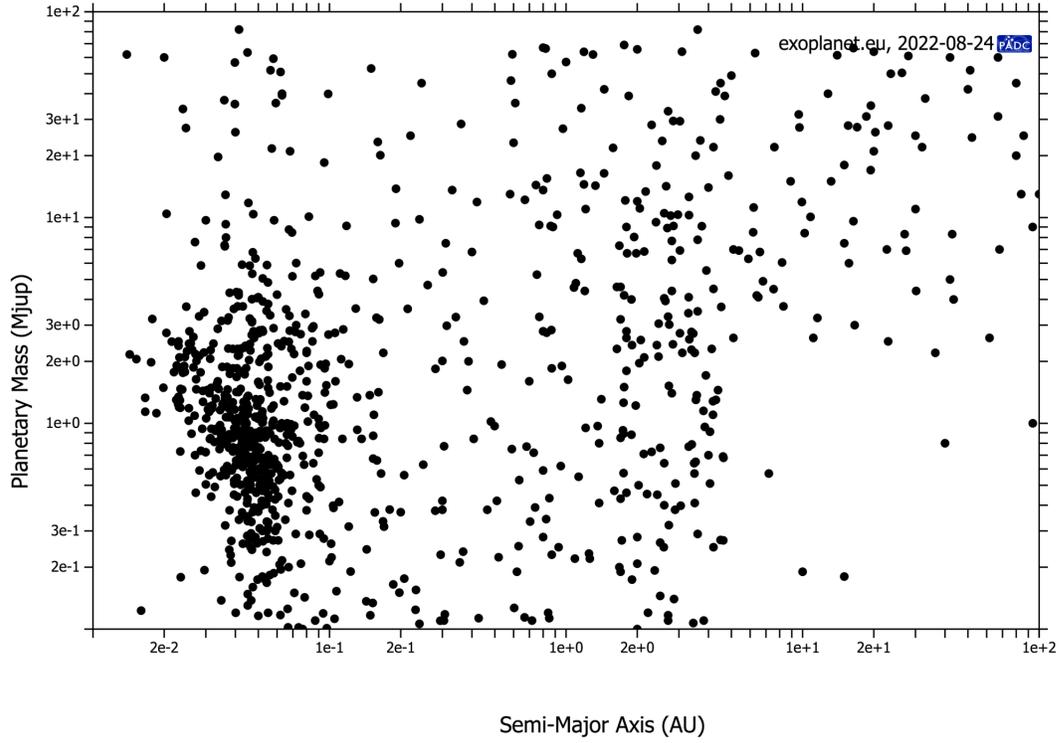}
\vspace{0.5in}
\caption{Exoplanet masses as a function of orbital semimajor axis from the Extrasolar
Planets Encyclopaedia (exoplanet.eu) as of 24 August 2022.}
\end{figure}
\clearpage

\begin{figure}
\vspace{-1.0in}
\includegraphics[scale=.80,angle=-90]{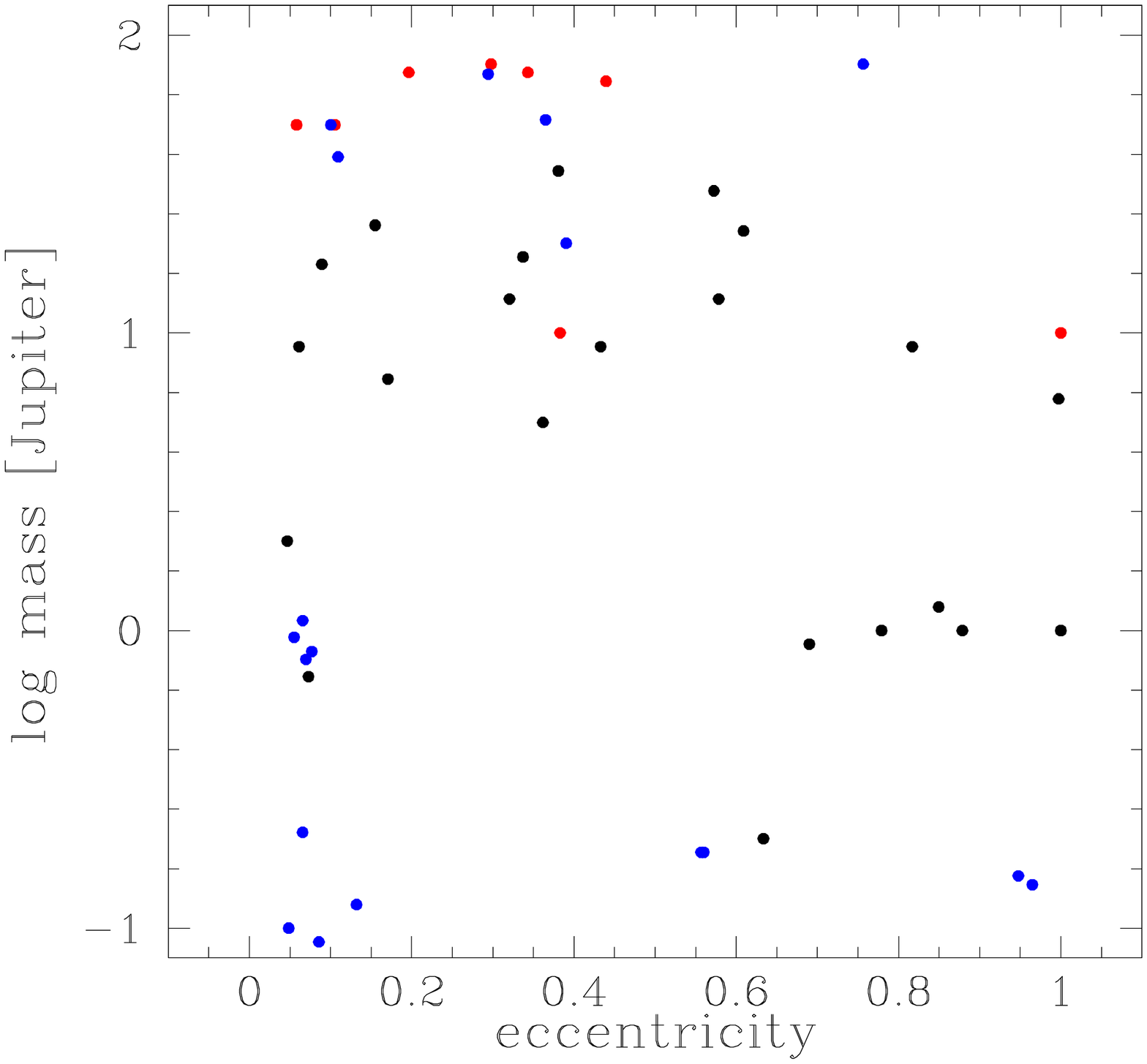}
\vspace{0.5in}
\caption{Sink particle masses as a function of the orbital eccentricity at the final times 
for the models. Black dots are for
models that started with 20 au radius disks, red dots are for 30 au disks, and blue
dots are for 60 au radius disks (see Table 1).}
\end{figure}
\clearpage

\begin{figure}
\vspace{-1.0in}
\includegraphics[scale=.60,angle=-90]{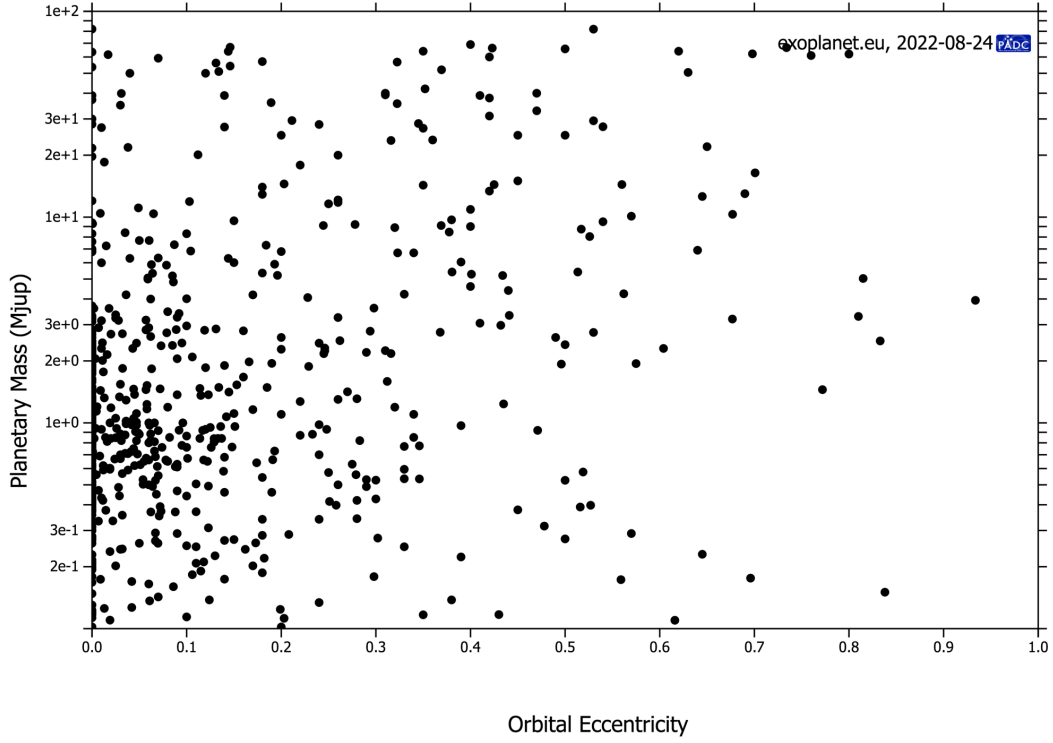}
\vspace{0.5in}
\caption{Exoplanet masses as a function of orbital eccentricity from the Extrasolar
Planets Encyclopaedia (exoplanet.eu) as of 24 August 2022.}
\end{figure}
\clearpage

\end{document}